\documentclass[reprint, amsmath,amssymb,aps,superscriptaddress]{revtex4-2}
\usepackage{graphicx,enumerate,verbatim,bbold}
\usepackage{amsmath,amssymb,amsthm,float,mathrsfs}
\usepackage{dsfont}
\usepackage{physics}
\usepackage[colorlinks]{hyperref}
\usepackage[T1]{fontenc}
\usepackage[utf8]{inputenc}
\usepackage{lmodern}
\usepackage{braket}
\usepackage{bm}
\usepackage{caption}
\usepackage{subcaption}
\usepackage{dcolumn}
\usepackage{hyperref}

\usepackage[usenames,dvipsnames]{color}

\newcommand{\unit}{1\!\!1}
\newcommand{\be}{\begin{equation}}
\newcommand{\ee}{\end{equation}}
\newcommand{\ba}{\begin{array}}
\newcommand{\ea}{\end{array}}
\newcommand{\bqa}{\begin{eqnarray}}
\newcommand{\eqa}{\end{eqnarray}}

\newcommand{\Imperial}{Blackett Laboratory, Imperial College London, SW7 2AZ, United Kingdom}
\newcommand{\ImperialMath}{Department of Mathematics, Imperial College London, London, SW7 2AZ, United Kingdom}
\newcommand{\HZDR}{Helmholtz-Zentrum Dresden-Rossendorf, Bautzner Landstraße 400, 01328 Dresden, Germany}

\begin{document}

\title{Generally noise-resilient quantum gates for trapped-ions}

\author{Modesto Orozco-Ruiz}\affiliation{\Imperial}
\author{Wasim Rehman}\affiliation{\ImperialMath}
\author{Florian Mintert}\affiliation{\Imperial}\affiliation{\HZDR}

\date{\today}
\begin{abstract}
We present an entangling gate scheme for trapped-ion chains that achieves high-fidelity operations with excited motional states despite multiple error sources. Our approach incorporates all relevant motional modes and exhibits enhanced robustness against both motional heating effects and detuning errors, critical features for building robust and scalable trapped-ion quantum computers.
\end{abstract}

\maketitle

\section{Introduction} \label{Introduction}

Trapped-ion quantum computing holds promise for building scalable and fault-tolerant quantum computers due to the exceptional control that can be exerted over individual qubits and their collective interactions \cite{ladd2010quantum}. This exquisite control enables the implementation of the two-qubit entangling gate -- a fundamental building block for quantum algorithms. Relying on the Coulomb interaction between ions, this gate utilizes the shared motion of the ions to generate entanglement in their electronic states \cite{bruzewicz2019trapped}. 

The M{\o}lmer-S{\o}rensen gate \cite{sorensen1999quantum} has emerged as one of the most promising implementations in trapped-ion quantum computing, with fidelities exceeding $99\%$ in both optical and hyperfine qubits \cite{gaebler2016high, harty2016high, weidt2016trapped, benhelm2008towards}. Exploiting precisely tuned bichromatic laser fields that couple internal and motional states, the M{\o}lmer-S{\o}rensen gate proves to be insensitive to the initial motional state, offering substantial advantages for scalability and error correction \cite{valahu2022quantum, haddadfarshi2016high, lishman2020trapped, webb2018resilient, hayes2012coherent}. 

However, despite robustness to initial motional state, current entangling gates remain vulnerable to various experimental imperfections, due to their inherent coupling with the environment  \cite{valahu2022quantum, haddadfarshi2016high, sorensen1999quantum}, as well as to systematic errors resulting from imperfect calibration of system parameters.
The complex interplay of error mechanisms in entangling gates presents a significant obstacle to the development of a robust scheme against multiple errors. While some approaches have demonstrably enhanced resilience against specific noise sources, optimizing resilience against one mechanism typically compromises resilience against another. For example, increasing laser power for faster gates can alleviate detuning errors but exacerbate off-resonant coupling to undesired carrier transitions \cite{randall2015high}. Similarly, suppressing motional mode cross-talk in many-ion systems by slowing down the gate operation can lead to reduced coherence time and imperfect coupling \cite{choi2014optimal, lu2019global}. These conflicting effects highlight the need for control protocols capable of simultaneously addressing diverse error sources.

Motional heating is arguably a primary source of errors in current implementations, with electric and magnetic field noise leading to unwanted phonon exchange that disrupts the motional states of the trapped ions \cite{randall2015high}. Similarly, frequency fluctuations of the vibrational modes induce dephasing, ultimately compromising the gate fidelity. To address these challenges, several schemes have been proposed to enhance the motional robustness of entangling gates by engineering optimal ion trajectories in phase space. One approach involves modulating sideband field parameters (phase, frequency or amplitude \cite{zarantonello2019robust, milne2020phase}) to create phase-space trajectories with minimal susceptibility to heating. Alternatively, a multi-tone driving strategy employs a combination of detuned fields with varying amplitudes for each ion sideband, achieving comparable robustness gains \cite{sameti2021strong}. However, despite their demonstrated success \cite{webb2018resilient}, these techniques are inherently limited by the constraints of the Lamb-Dicke regime within which entangling gates operate \cite{sameti2021strong}. This regime, characterized by weak laser-ion interactions, requires low laser power for high-fidelity operations, leading to slower gate times. Additionally, shared motional states must be cooled close to their ground state, posing a scalability bottleneck for proposals involving active ion shuttling \cite{lekitsch2017blueprint, orozco2023optimal}, as such movements inherently excite these states. Deviations from these conditions, such as using stronger laser fields or excited motional states, result in compromised fidelities. 

Logic gates beyond the Lamb-Dicke regime offer significant improvements, including high fidelities with high motional states and significantly faster quantum operations. One approach employs a series of complex pulses inducing state-dependent kicks exceeding the motional frequencies \cite{wong2017demonstration}. While this method demonstrably enhances speed, current fidelities remain moderate due to experimental complexity and unwanted coupling to additional motional modes. 

An alternative strategy employs simpler driving fields and higher-order sideband transitions that are designed to cancel undesired contributions from the non-linear spin-motion coupling beyond the Lamb-Dicke regime \cite{sameti2021strong}. This approach holds promise for achieving high fidelities in the strong-coupling regime, but relying on driving more sideband transtions is more vulnerable to motional heating. 

Another significant challenge faced by current entangling gates involves scaling to larger ion chains with a complex multi-mode structure \cite{haffner2005scalable, kim2009entanglement, landsman2019two}. This not only amplifies noise sensitivity due to increased crosstalk between ions \cite{lu2019global} but also significantly complicates control protocols, specially far from the Lamb-Dicke regime. This occurs because the coupling strength corresponding to each motional mode becomes comparable in magnitude and it varies across different ions in the chain. Selectively driving a single mode in such systems inevitably triggers higher-order phonon exchange processes, unintentionally coupling non-targeted spectator modes to the qubit states. This presents a significant challenge in achieving high-fidelity gates, requiring control strategies distinct from those employed in paradigms with fewer modes, where precise control over individual modes is sufficient.

This manuscript introduces a gate scheme that integrates multiple strategies to combat diverse error sources in trapped-ion quantum computing. Our control framework achieves high-fidelity entanglement in hot trapped-ions within a multi-mode system, demonstrating inherent resilience against motional heating and increased robustness against fluctuations in both qubit frequency and normal-mode frequencies.

\section{Hamiltonian model} \label{hamiltonian_model}

The dynamics of a system comprised of the internal degrees of freedom of $N$ ions and $M$ collective motional degrees of freedom is induced by the Hamiltonian
\begin{align} \label{eq:hamiltonian_0}
H_0=
\sum_{j=1}^{N}\frac{\omega_j}{2}\sigma_z^{(j)}+
\sum_{l=1}^{M}\nu_l a_l^\dagger a_l\ ,
\end{align}
where $\sigma_z^{(j)}$ is the Pauli-Z operator for ion $j$ with resonance frequency $\omega_j$, and $a_l^\dagger (a_l)$ are the creation (annihilation) operators for the collective motional degree of freedom $l$ with resonant frequency $\nu_l = \kappa_l \nu$ in terms of the trapping frequency $\nu$.

The interaction with an external light field, such as a laser field, can excite both internal and motional degrees of freedom of the ions.
The respective interaction Hamiltonian reads
\begin{align}
H(t) &= \sum_{j=1}^{N}f_j(t) \sigma_+^{(j)}
\prod_{l=1}^M e^{i\eta_{jl} \left( a_l+ a_l^{\dagger}\right)}+ h.c.\ , 
\label{eq:original_H}
\end{align}
in terms of the raising operator $\sigma_+^{(j)}$ of the internal degree of freedom of ion $j$, and the driving functions $f_j(t)$ that encapsulate the generally time-dependent Rabi-frequency and a time-dependence due to carrier frequencies of the light field.
The Lamb-Dicke parameters $\eta_{jl} = \chi_{jl}\Lambda$ characterize the coupling between the state transition in ion $j$ and each motional state transition in mode $l$ due to light field interaction. They depend on the expansion coefficients $\chi_{jl}$ and the coupling factor $\Lambda=\frac{2\pi}{\lambda}\sqrt{\frac{\hbar}{2 m \nu}}$ defined in terms of the wavelength of the light field $\lambda$ and the mass $m$.

Since entangling gates are realized via the exchange of virtual phonons in the collective motion of the ions, it is instructive to express the interaction Hamiltonian as
\begin{align}
H(t) &=  \sum_{j=1}^{N} \tilde f_j(t) \sigma_+^{(j)}
\prod_{l=1}^M \sum_{k=-\infty}^{\infty} \mathcal D_{l,k} (\eta_{jl})+ h.c.\ , 
\label{hamiltonian_modes}
\end{align}
in terms of the operators
\begin{align}
\label{eq:displacement}
\mathcal D_{l,k} (\eta) =\sum_{n=0}^{\infty} \left(i \eta \right)^{2n + k} \frac{a_l^{\dagger n+k}a_l^n}{(n+k)!n!}\ ,
\end{align}
and
\begin{align}
\label{eq:displacement2}
\mathcal D_{l, -k}(\eta) =  (-1)^{k} \mathcal D_{l, k}^{\dagger}(\eta)\ ,
\end{align}
for $k\geq 0$,
that encapsulate all processes resulting in the creation (of $k>0$) and annihilation ($k<0$) of $k$ phonons in a given mode,
and a renormalised driving function
\begin{align} \label{eq:renormalised}
\tilde f_j (t) = f_j(t)\ e^{-\frac{1}{2}\sum_{l=1}^M \eta_{jl}^2}\ .
\end{align}

The operators $\mathcal D_{l,k}$ capture the $k$-th order sideband transitions of the motional mode $l$.
A generic driving pattern $\tilde f_j (t)$ can have spectral components that are close to resonance with each of these (red and blue) sideband transitions.
In standard rotating wave approximation, each spectral component needs to be considered in the Hamiltonian only in close-to-resonant terms, but it can be neglected in any far-off-resonant term. 
A general parametrization of driving patterns is thus given by
\begin{align} \label{eq:driving_patterns}
\tilde f_j= -i\sum_{l}\frac{1}{\eta_{jl}} \sum_k F_{l,k}^{(j)}(t)\ e^{-i (k \nu_l+\omega_j) t}\ ,
\end{align}
where the factors $e^{-i (k \nu_l+\omega_j) t}$ capture the time-dependence necessary for resonance with a given sideband transition, as demonstrated by expressing $H(t)$ (Eq.~\eqref{hamiltonian_modes}) in the interaction picture with respect the free Hamiltonian $H_0(t)$ (Eq.~\eqref{eq:hamiltonian_0}). The factors $F_{l,k}^{(j)}(t)$  have a slow time-dependence compared to $\nu_l$ and account for finite detuning and temporal modulation that can be used to achieve the desired robustness. Finally, the factor $\eta_{jl}^{-1}$ reflects the fact that weaker interaction (smaller $\eta_{jl}$) requires proportionally stronger driving fields to maintain constant interaction strength.

A coherent gate requires that there are no changes to the motional state that depend on the internal state of the ions, and achieving this is facilitated in terms of driving patterns in which the driving amplitude $F_{l,-k}^{(j)}$ for any blue sideband transition ($k>0$) is given by the driving amplitude $F_{l,k}^{(j)}$ for the corresponding red sideband transition following the relation
\begin{align}
F_{l,-k}^{(j)}=(-1)^k\left(F_{l,k}^{(j)}\right)^\ast\ ,
\end{align}
where the factor $(-1)^k$ has its origin in Eq.~\eqref{eq:displacement2}.

With any such driving pattern, the interaction Hamiltonian reduces to
\begin{align}
\label{eq:hamiltonian_simplified}
    H(t) = \sum_{j=1}^N \sigma_y^{(j)} \sum_{l,k>0} \frac{F_{l,k}^{(j)}(t)}{\eta_{jl}}\ \mathcal D_{l,k} \prod_{l'\neq l} \mathcal D_{l', 0} + h.c.\ ,
\end{align}
with $\sigma_y = i(\sigma_-  - \sigma_+)$, and with factors
\begin{align}
\mathcal D_{l,0}=\sum_{n=0}^{\infty} \frac{\left(i \eta \right)^{2n}}{n!^2} a_l^{\dagger n} a_l^n
\end{align}
that describe phonon-conserving, but phonon-number-dependent energy shifts.

\section{Driving patterns} 

A light field modulated as described in Eq.~\eqref{eq:driving_patterns}  allows the engineering of a time-dependent Hamiltonian (Eq.~\eqref{eq:hamiltonian_simplified}) with desired dynamics. By carefully choosing the driving functions $F_{l,k}^{(j)}(t)$, the interaction with ions can be tailored to induce specific processes necessary for entangling state generation.

Crucially, not all motional modes of the ion chain are required for entanglement generation. Only a subset of these modes, along with a finite number of their sideband transitions, play a crucial role. The M{\o}lmer-S{\o}rensen~(MS) entangling gate~\cite{sorensen1999quantum, sorensen2000entanglement} exemplifies this principle, where entanglement is achieved by driving solely the first sideband of a chosen set of motional modes. 

This specific solution to Eq.~\eqref{eq:hamiltonian_simplified} requires the weak coupling regime or motional states near their ground states as this allows approximating the displacement operators for the driven modes $d$ in Eq.~\eqref{eq:hamiltonian_simplified} to the lowest order in $\eta$ as $\mathcal{D}_{d,1} \approx i\eta_{d} a_{d}^{\dagger}$, while those for the spectator modes reduce to the identity $\mathcal{D}_{l',0} \approx \unit$. This first-order simplification implies that spectator modes do not couple with the driven modes (see Eq.~\eqref{eq:hamiltonian_simplified}) and can therefore be safely neglected,  thus resulting in the simplified Hamiltonian $H_{MS}(t) = i S_y \sum_d \left( F_d(t) a_d^{\dagger} - F_d^*(t) a_d \right)$, with the collective spin operator $S_y = \sum_j \sigma_y^{(j)}$. With this simplified Hamiltonian, a carefully chosen set of driving functions $F_d(t)$ enables the implementation of the entangling gate $U = e^{i \phi_T \sum_{j\neq k}\sigma_y^{(j)}\sigma_y^{(k)}}$ acting on all ion pairs.

\subsection{Non-identical ion species and multi-mode systems}

Although the Hamiltonian $H_{MS}$ offers a straightforward approach for creating entangled states, it comes with certain limitations. While a global driving field ($f_{j}(t)=F(t)$ for all ions $j$) works for identical ion species, two key challenges arise when dealing with non-identical ions. Firstly, their differing resonance frequencies $\omega_j$  lead to the acquisition of distinct phase factors $e^{i \omega_j t}$ during their dynamics. Secondly, non-identical ions couple to the mediating modes with varying strengths $|\eta_{j,l}|$. This dependence on the specific ion $j$ results in different scaling factors $e^{\frac{1}{2} \eta_{j, l}^2}$ arising in the expansion of the interaction Hamiltonian in Eq.~\eqref{eq:original_H}. 

These challenges related to non-identical ions can be addressed using a more sophisticated individual addressing approach instead of a single, global field. However, a more fundamental limitation lies with the Lamb-Dicke approximation used to obtain the simplified first-order Hamiltonian $H_{MS}$. This approximation restricts the applicability of $H_{MS}$ to systems operating within the weak coupling regime or with motional states close to their ground states, as dictated by the inequality $\frac{\eta_l^2}{2}n_l \ll 1$ \cite{sorensen2000entanglement, sameti2021strong}. Systems that do not satisfy this constraint necessitate alternative approaches to obtain high-fidelity operations.

\subsection{Beyond the Lamb-Dicke approximation}

Moving beyond the Lamb-Dicke approximation, either due to stronger ion-motion interactions or higher motional states, necessitates considering higher-order phonon-exchange processes within the displacement operators     Eq.~\eqref{eq:displacement}. This leads to more complex spin-motion interactions for driven modes ($\mathcal{D}_{l,k} \approx \frac{(i\eta)^k}{k!} a_l^{\dagger k} + \frac{(i\eta)^{k+2}}{(k+1)!} a_l^{\dagger (k+1)} a_l + \cdots$) and, crucially, indirect coupling between driven and spectator modes via the non-trivial products $\mathcal D_{l,k}\prod_{l'\neq l} \mathcal (\unit - \eta_{l'}^2 a_{l'}^{\dagger} a_{l'} + \cdots)$ in Eq.~\eqref{eq:hamiltonian_simplified}.

Consequently, control schemes beyond the Lamb-Dicke regime must account not only for higher-order phonon processes in the driven modes but also for the phonon exchange processes in the remaining non-addressed collective vibrational modes. To illustrate this phenomenon, consider a two-ion system where driving the first sideband of the center-of-mass mode induces phonon exchanges like $(\eta_{1}^3/2) a_{1}^{\dagger 2} a_{1}$ beyond the first-order term $\eta_{1} a_{1}$. Similarly, even though the stretch mode is not directly addressed, it couples through this driving and contributes to dynamics with processes like $\eta_{2}^2 a_2^{\dagger} a_2$.

Therefore, capturing these inter-mode dependencies and designing high-fidelity entangling gates for many-mode ion chains beyond the Lamb-Dicke regime necessitates a higher-order expansion of the interaction Hamiltonian $H(t)$. This expansion reveals exchange processes neglected in the standard treatment, requiring the development of a new control strategy that eliminates their unwanted contributions at the gate time. 

Following this crucial first step, designing a robust two-qubit entangling gate involves selecting a suitable set of driven motional modes $L$ and the specific sidebands for each of those modes $K(l)$. These choices determine which processes are driven resonantly. The temporal modulation of the driving functions, $F_{l,k}^{(j)}$ is then tailored to achieve entanglement at the desired gate time $T$ while actively suppressing the unwanted processes and maximizing the gate's overall robustness.

\subsection{Driven sidebands}

The specific construction presented here enables the implementation of an entangling gate in a multi-mode system of $N$ trapped ions with high motional states and offers enhanced resilience against both motional heating and combined errors in qubit and vibrational frequencies. While alternative choices exist, the proposed configuration minimizes the number of driven sidebands required for achieving the desired robustness, offering a practical advantage for experimental implementation.

In a system with $M$ motional modes, the proposed gate implementation involves simultaneous driving of all modes ($L=M$). The first-order sideband ($k=1$) of a designated central mode, typically chosen for its stronger spin-motion coupling, acts as the primary bus mode of the entangling process. Meanwhile, other modes (together with the central one) contribute through second-order sidebands ($k=2$) to actively suppress unwanted contributions arising from higher-order terms in the Hamiltonian. 

With such choice, the Hamiltonian Eq.~\eqref{eq:hamiltonian_simplified} simplifies to

\begin{align}
\label{eq:hamiltonian_simplified2}
H_c(t) &=  \sum_{j=1}^N \sigma_y^{(j)} 
\left(
\frac{F_{1,1}^{(j)}}{\eta_{j1}} \mathcal D_{1,1} D_1 + \sum_{l=1}^M \frac{F_{l,2}^{(j)}}{\eta_{jl}} \mathcal D_{l,2} D_l\right)  + h.c.,
\end{align}

with $D_k=\prod_{l'\neq k} \mathcal D_{l', 0}$ being the product of phonon-conserving operators in the spectator modes.

Building upon Eq.~\eqref{eq:hamiltonian_simplified2}, one can derive time-dependent functions $F_{1,1}^{(j)}$ and $F_{l,2}^{(j)}$ specifically to achieve an exact solution of the driven dynamics up to a particular order of $\eta$. Although higher-order solutions are theoretically feasible, the present solution results from an expansion to order $\order{\eta^3}$, significantly enhancing first-order expansions while minimizing the number of sidebands required for achieving the desired robustness. With their detailed derivation available in Sec.~\ref{methods}, these functions are given by

\begin{subequations}
\label{eq:parametrisation}
\begin{alignat}{2}
\label{eq:parametrisation_a}
F_{1,1}^{(1)} &= &&\, F_{1,1}^{(2)} = \Omega \left( e^{2i \delta t}  - \frac{3}{2} e^{3i \delta t}\right) \\
\label{eq:parametrisation_b}
F_{l,2}^{(1)} &= &&\, s_{1l}\Omega  \frac{\tilde \eta_l}{\eta_{1l}} e^{i \delta t} \\
\label{eq:parametrisation_c}
F_{l,2}^{(2)} &= &&\, s_{1l} \Omega \frac{\tilde \eta_l  }{ \eta_{2l}} e^{i \delta t}
\end{alignat}
\end{subequations}

where $l = 1, \cdots, M$ refers to the motional mode, $s_{jl}=\mbox{sign}(\eta_{jl})$ and $\tilde \eta_l := \frac{\sqrt{5}}{2} \sqrt{\eta_{1l}^2 + \eta_{2l}^2}$. 

\begin{figure}[t]
\centering
\includegraphics[width=0.499\textwidth]{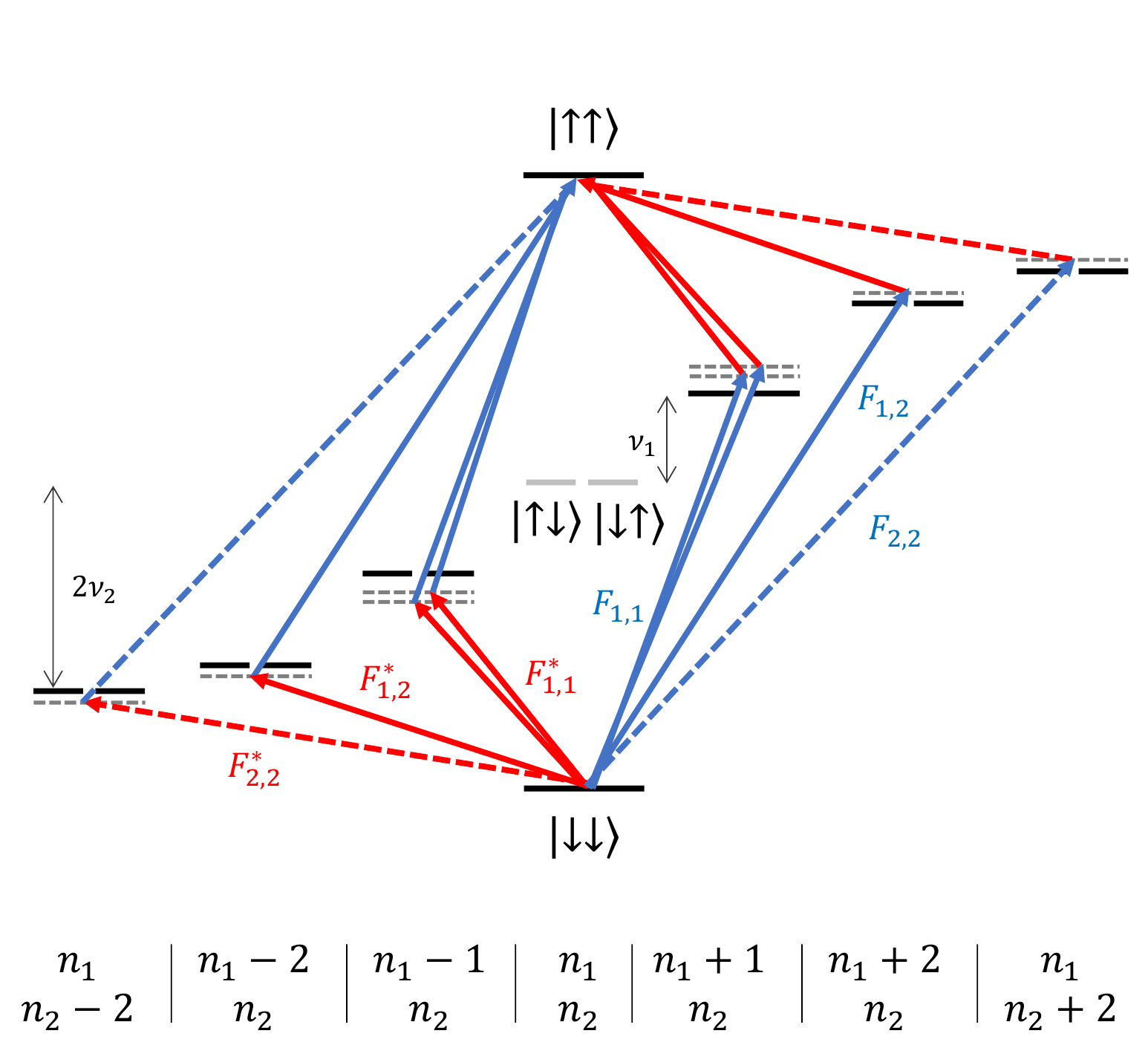}
\caption{Energy level diagram for two identical ions and two motional modes. Note that only energy levels involved in the driving process are shown for simplicity. Solid blue (red) sidebands represent transitions with a phonon gain (loss) in motional mode 1. Dashed blue (red) sidebands represent similar processes for motional mode 2.}
\label{fig:1}
\end{figure}

Consider a specific pair of ions selected for entanglement generation within the $N$-ion chain. According to Eq.~\eqref{eq:parametrisation_a}, both ions receive identical bichromatic modulation with detunings $2\delta$ and $3\delta$ ions targeting the first sideband of the primary motional mode (see Fig.~\ref{fig:1} for the case of two identical ions). With such a modulation, resilience against motional heating is enhanced, while the specific detuning ratio optimizes gate speed. Eqs.~\eqref{eq:parametrisation_b} and \eqref{eq:parametrisation_c}, in turn, specify different drivings for each ion in the pair targeting second order sidebands of all vibrational modes (see Fig.~\ref{fig:1}). These suppress unwanted processes that emerge as higher-order (in $\eta$) contributions to the system's dynamics and include mode-dependent amplitudes to compensate for the different coupling strengths across the different modes. Finally, the Rabi amplitude $\Omega$ is specified in terms of the detuning $\delta$ as a solution to the equation

\begin{align} \label{eq:entangling}
 -6 \frac{\Omega^4}{\delta^4} \left( \sum_j \eta_{j1}^2 \right) +  \frac{\Omega^2}{\delta^2} \left(  2 +  \sum_{l, j} \eta_{jl}^2 \right)  = \frac{2\phi_T}{5\pi} \ ,
 \end{align}
 
defined in terms of the entangling phase $\phi_T$ in the unitary gate $U$. In the weak-coupling regime, where $\eta_{jl} \approx 0$, the higher-order terms in the equation become negligible and the entangling condition reduces to the standard MS case with a bichromatic driving of the form Eq.~\eqref{eq:parametrisation_b}, \textit{i.e.}, $\Omega^2/\delta^2 = \phi_T/ 5\pi$.

\begin{figure*}[t]
     \centering
     \begin{subfigure}[b]{0.49\textwidth}
         \centering         \includegraphics[width=\textwidth]{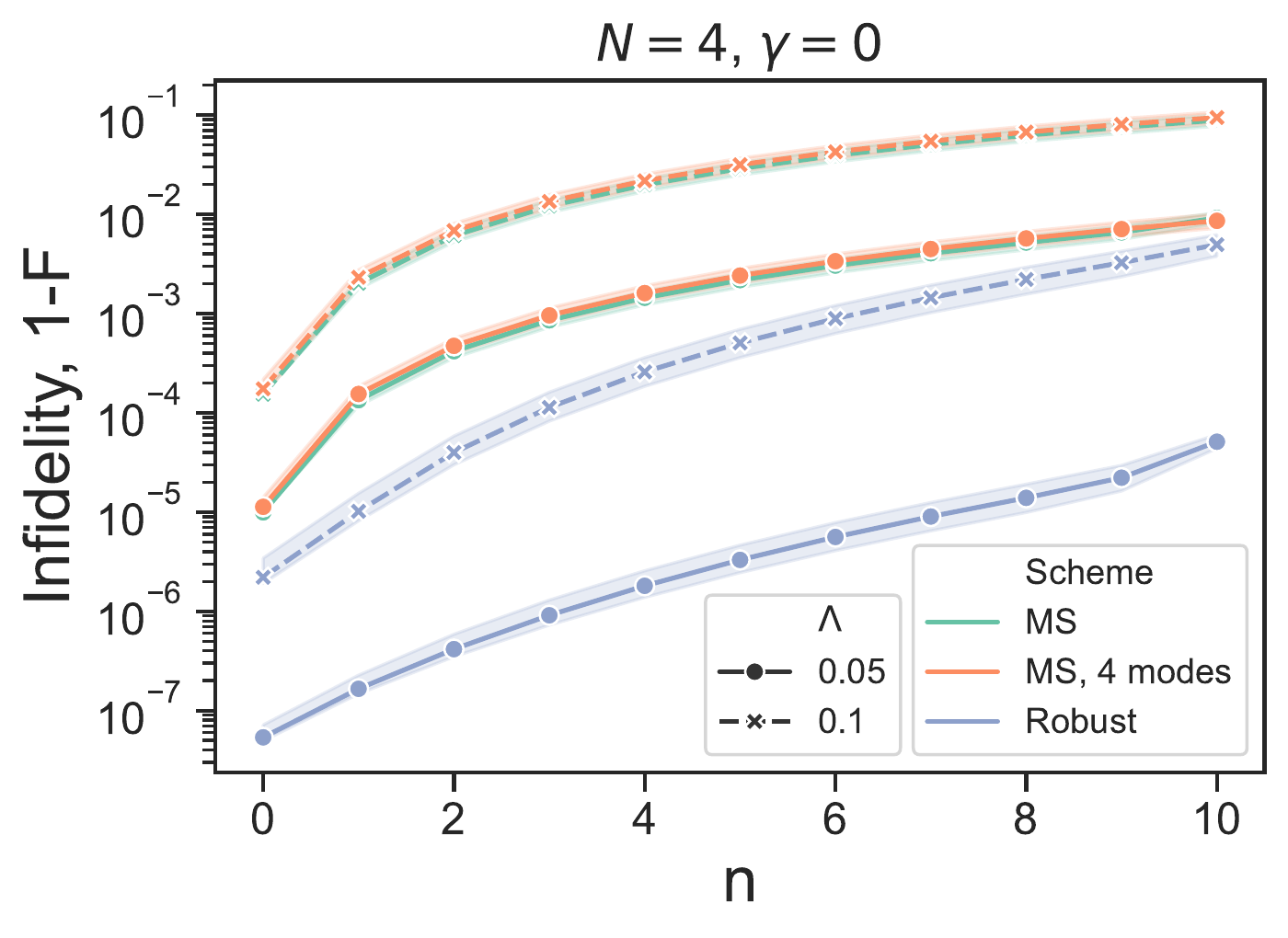}
         \caption{}
         \label{fig:2a}
     \end{subfigure}
    \begin{subfigure}[b]{0.49\textwidth}
         \centering
         \includegraphics[width=\textwidth]{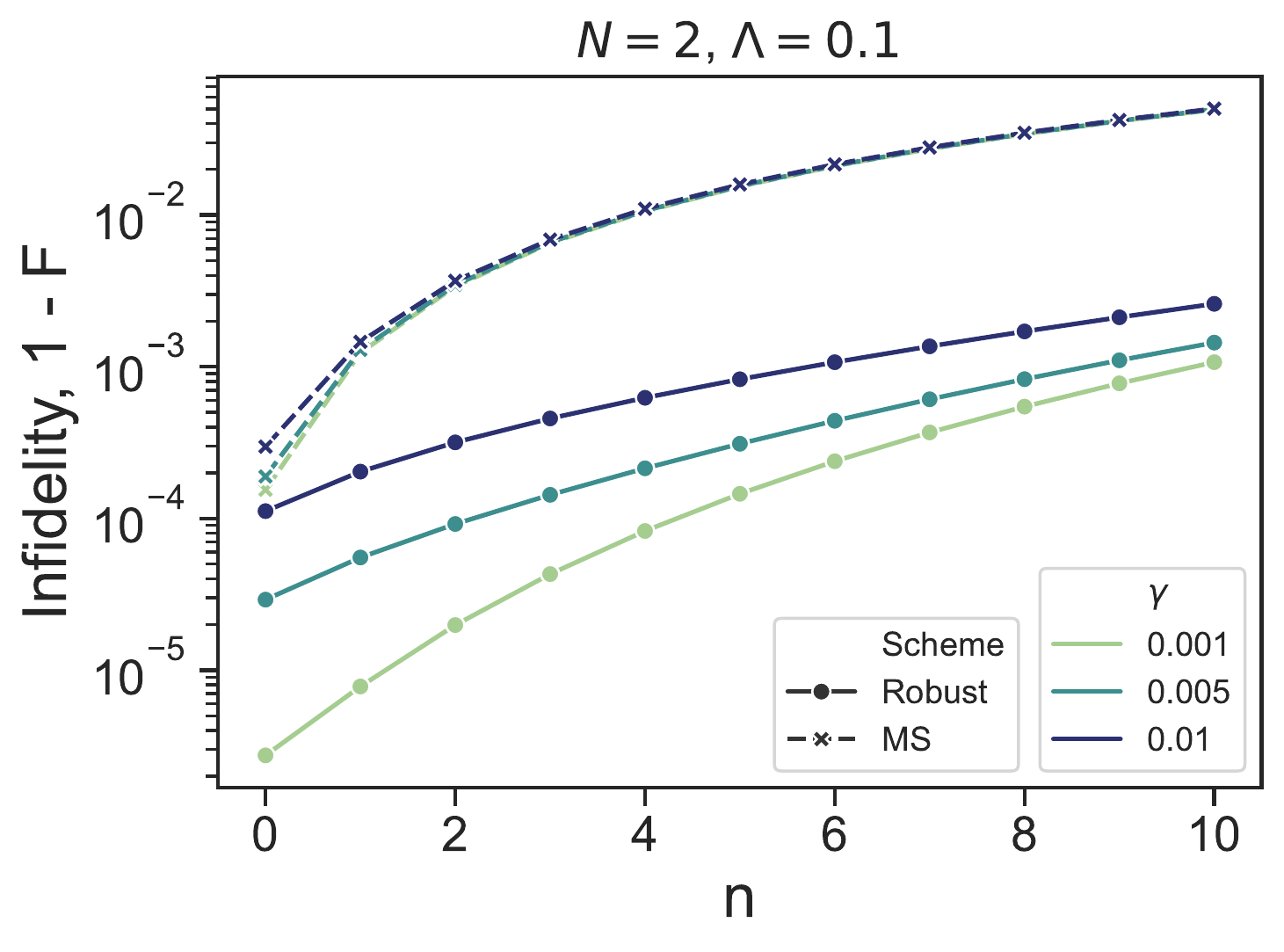}
         \caption{}
         \label{fig:2b}
     \end{subfigure}
     \hfill
     \caption{Infidelity $1 - F$ of the entangling gate as a function of the initial Fock state occupation $n$ for a 4-qubit (left panel) and 2-qubit (right panel) systems.  In the left panel, each data point represents the average infidelity across all qubit pairs, with the shaded area showing the range of values found across different pairs. The infidelity corresponds to the standard MS gate (green), the MS gate applied to all four modes (orange), and the present gate (blue). No motional heating ($\gamma = 0$) is assumed and different values of the coupling strength $\Lambda$ are shown with different markers. In the right panel, the impact of motional heating is evaluated for a fixed spin-motion coupling $\Lambda=0.1$. Decay rates are represented by colors ranging from light green (low) to purple (high) and marker styles differentiate the standard MS gate (crosses) from the proposed robust scheme (circles). In all simulations, the amplitude of the driving fields in the present scheme is limited to match the amplitude of the MS gate, ensuring a fair comparison.}
    \label{fig:2}
\end{figure*}

\section{Results}

The proposed control strategy enables the implementation of high-fidelity entangling gates for trapped-ions in noisy environments, even when the ions share multiple collective vibrational modes and are in high motional states. Gate performance can be quantified using the gate fidelity, which is given by
\begin{align}
\label{eq:gate_fidelity}
    F(\rho_M) = \frac{1}{d^2} \sum_{\alpha,\beta=1}^{d} \mel{\alpha}{U^{\dagger} \Tr_M\left(\mathcal{E}(\dyad{\alpha}{\beta}\otimes \rho_M)\right) U}{\beta}, 
\end{align}
where the states $\ket{\alpha}, \ket{\beta}$ span the $d$-dimensional Hilbert space of spin states (with $d=2^N$), $\rho_M$ is the initial density matrix of motional states, $U$ represents the target unitary and $\mathcal{E}(\cdot)$ the actual (potentially non-unitary) channel corresponding to the implemented gate. 
Evaluating the fidelity $F(\rho_{th})$ for thermal states ($\rho_{th} = \sum_{{\bf n}} P({\bf n}) \dyad{{\bf n}}$, with $\ket{{\bf n}} = \ket{n_1, n_2, \cdots n_M}$) is crucial, as these states are commonly encountered in real experiments. By exploiting the linearity of the trace, one can avoid the construction of the density matrix and instead compute the fidelity for a sufficiently large set of Fock states. Averaging these individual fidelities yields the thermal state fidelity $F(\rho_{th})= \sum_{{\bf n}} P({\bf n}) F(\dyad{{\bf n}})$.

The following discussion delves into the dependence of the fidelity $F$ on several key parameters relevant to an experimental implementation, including chain size $N$, initial motional occupations $\ket{{\bf n}}$, and the impact of non-idealities like motional heating and frequency errors in the vibrational modes $\epsilon_{\nu}$ and spin transition frequencies $\epsilon_{\omega}$.

\begin{figure*}[t]
     \centering       \includegraphics[width=0.99\textwidth]{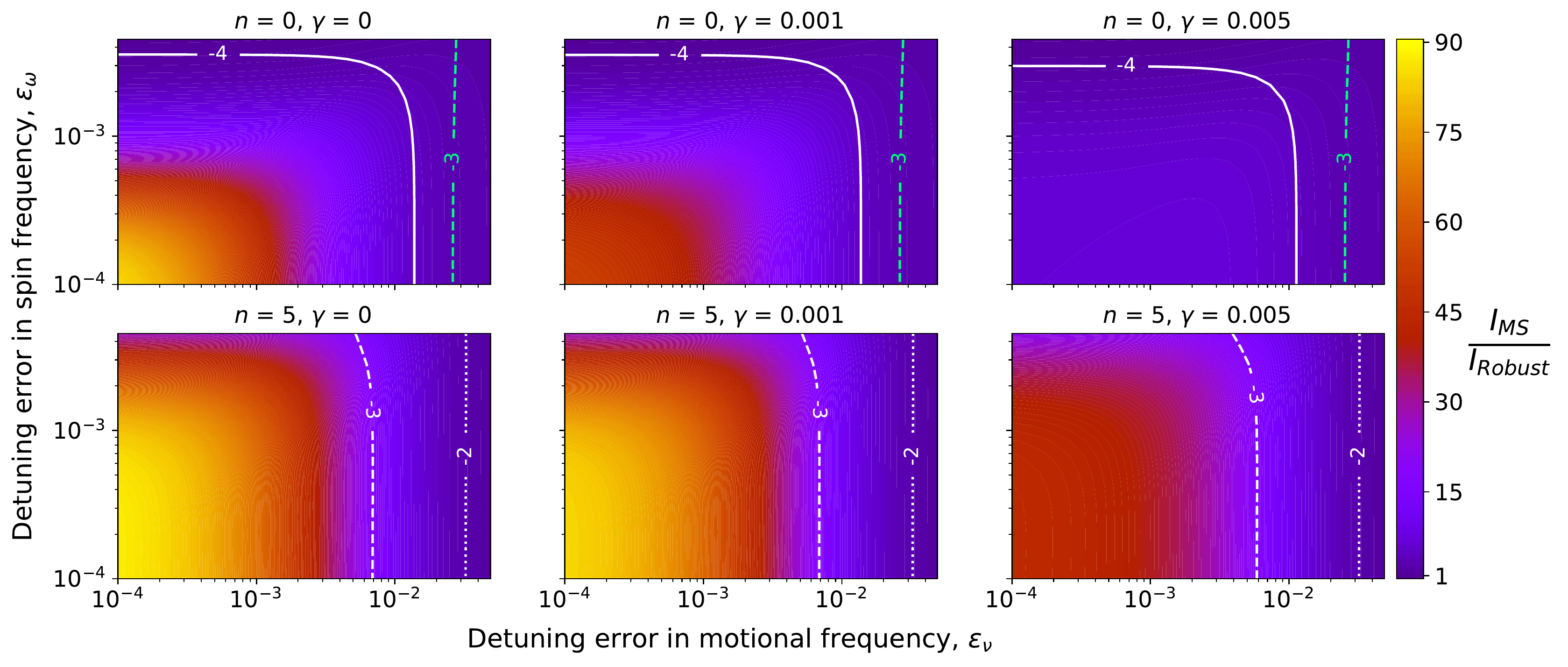}
     \label{fig:3a}
     \caption{Combined impact of vibrational and spin frequency errors on the performance of the present entangling scheme (Robust) compared to the standard MS gate in a two-ion system. Each panel shows the ratio of MS gate infidelity to SC gate infidelity $I_{MS}/I_{Robust}$ for a specific initial motional state $\ket{n}$ and decay rate $\gamma$, with fixed coupling strength $\Lambda=0.1$. The $x$-axis represents (in log scale) errors in vibrational frequency $\epsilon_{\nu}$; while the $y$-axis depicts symmetric errors in spin frequencies $\epsilon_{\omega}$. Contour lines reveal actual fidelities achievable by each method. Solid lines enclose regions where infidelities remain below $10^{-4}$, dashed lines correspond to $10^{-3}$ and dotted lines to $10^{-2}$. Green lines represent the MS gate, while white ones represent the robust scheme.}
    \label{fig:3}
\end{figure*}

\subsection{Multi-mode systems with no dissipation or frequency errors}

Unlike simpler two-ion systems, designing entangling gates for larger, multi-mode, systems requires accounting for variations in coupling strengths $\eta_{jl}$ across ion pairs. This key challenge is directly addressed by the proposed control scheme. Fig.~\ref{fig:2a} exemplifies this point by depicting the infidelity $1-F$ of the present gate in a 4-ion system under ideal conditions (no dissipation or frequency errors). The x-axis represents the initial motional Fock state $\ket{n}^{\otimes 4}$,  and the shaded area shows the range of infidelities across all ion pairs, with marker points indicating the average value. Colors distinguish the present gate (in blue), the standard MS gate applied to the center-off-mass mode (in green), and the MS gate applied to all four motional modes~\cite{sackett2000experimental} (in orange); while different marker styles represent varying coupling strengths $\Lambda$. 

Crucially, under ideal conditions, the present gate consistently outperforms both MS variants across all initial states and coupling strengths. This advantage is particularly evident in weak-coupling regimes ($\Lambda=0.05$, circles). Even with high motional states ($n=10$), it achieves infidelities below $10^{-4}$, a remarkable two orders of magnitude lower than any MS variant. Notably, this performance is independent of the specific ion pair within the four-ion system, demonstrating its robustness across various configurations, a crucial feature for practical applications. 

While the current analysis is limited to Fock states with $n = 10$ for a four-ion system, numerical evidence suggests this trend persists for significantly higher levels of motional excitation, indicating the gate's high performance even under substantial thermal noise. Indeed, fidelities exceeding $99\%$ are still achievable even at mean phonon occupations as high as $\bar n \sim 100$, as shown in Supp. Mat.~\ref{Supmat2} for a two-ion system.

\subsection{Motional heating}
The previous results suggest that the present entangling protocol allows high-fidelity entanglement without near-ground-state motional cooling, particularly advantageous for systems with weak spin-phonon coupling. However, a critical consideration is whether this benefit comes at the cost of increased susceptibility to motional decoherence due to the additional side-band driving. Crucially though, the temporal modulation of the driving fields, as defined in Eq.~\eqref{eq:parametrisation}, strategically mitigates heating effects, thereby enabling robust entanglement generation even in the presence of thermal noise. This protection, as detailed in Sec.~\ref{methods}, stems from the bichromatic driving in $F_{1,1}^{(1)}$ and $F_{1,1}^{(2)}$, which minimizes the impact of incoherent processes arising from motional heating to first order in $\eta$. Further enhanced robustness can be achieved by implementing a polychromatic version for all drivings in Eq.~\eqref{eq:parametrisation}, although at the cost of a more intricate driving scheme. 

Fig.~\ref{fig:2b} illustrates the performance of the proposed gate in the presence of motional heating for a two-ion system of identical species. In contrast to the prior scenario, the coupling strength remains uniform across all motional modes and ions (\textit{i.e.} $|\eta_{jl}|$ is independent of $j$ and $l$) and, therefore, local drivings can be relaxed to be global. The infidelity $1-F$ is represented as a function of the motional occupation for a fixed coupling strength $\Lambda = 0.1$ and different heating decay rates $\gamma$ (see Eq.~\eqref{eq:lindblad}).
The crosses correspond to the performance of the MS gate applied to the center-off-mass mode, while circles represent solutions derived from the parametrization in Eq.~\eqref{eq:parametrisation}. The results demonstrate a significant performance advantage of the present scheme over the MS approach across all decay rates, particularly in low-dissipative environments.
This enhanced robustness achieved with bichromatic driving enables fidelities close to $10^{-3}$ for motional occupations of up to $n=10$ in low and medium dissipative environments ($\gamma \leq 0.01$), compared to an infidelity close to $10^{-1}$ for the MS gate.


\subsection{Detuning errors}

Even minimal errors in vibrational and spin frequencies can significantly affect the gate fidelity. These errors, typically small compared to the resonant frequencies  (\textit{i.e.}, $\epsilon \ll \omega$), can be modeled as shifts of the form $\omega \rightarrow \omega + \epsilon$. Specifically, errors in motional frequencies manifest as shifts for each motional mode $l$, $\nu_l \rightarrow \nu_l + \epsilon_{\nu_l}$, whereas (symmetric) spin frequency errors are modeled as $\omega_j \rightarrow \omega_j + \epsilon_{\omega}$, affecting the resonant frequencies of the spin transitions. These small frequency drifts in vibrational and spin transitions lead to a crucial mismatch between the actual transition frequencies and the driving frequencies defined in Eq.~\eqref{eq:driving_patterns}, introducing an undesired detuning error that modifies the actual dynamics and may lead to potential performance degradation. 

Crucially, the properties that enhance resilience to motional heating (mainly closed phase-space trajectories and vanishing time-integral, as detailed in Supp. Mat.~\ref{Supmat1}) are also relevant for mitigating the impact of motional detuning errors \cite{valahu2022quantum}. Therefore, the proposed gate exhibits robustness against these errors by design. On the other hand, while not explicitly designed to address spin frequency detuning errors, the proposed gate still exhibits a relevant degree of robustness to them, as shown in Fig.~\ref{fig:3}.

Fig.~\ref{fig:3} depicts the combined impact of detuning errors in vibrational and spin frequencies for a two-ion system also experiencing motional heating. Each panel represents 
the ratio in infidelities between the MS gate (applied to the center-off-mass mode) and the present robust protocol, $I_{MS}/I_{Robust}$, using a color code ranging from purple (similar performance) to yellow (significant improvement). The panels vary across different initial motional states $\ket{\mathbf{n}}$ and decay rates $\gamma$; and the $x$ and $y$ axes correspond to errors in the vibrational $\epsilon_{\nu}$ (assumed to be the same across all modes for simplicity) and spin frequencies $\epsilon_{\omega}$, respectively.

The first row of panels shows the infidelity ratio for the initial motional ground state ($n=0$) with increasing decay rates from left to right. Here, the present scheme (robust) exhibits substantial improvement over the MS gate for small and moderate errors in motional frequencies
($ \epsilon_{\nu} < 0.5\%$), even with spin frequency shifts of up to $ \epsilon_{\nu} < 0.1\%$. Notably, reductions in infidelity of up to two orders of magnitude are achieved in the case with no motional heating and small frequency errors (yellow region in leftmost panel).
While this advantage diminishes with increasing motional heating (moving right within the first row), the present scheme maintains superior performance across all decay rates depicted, within the studied range of frequency errors.

Overall, when restricting motional frequency errors to a maximum of $1.5\%$, the proposed gate achieves fidelities exceeding $99.99\%$ (as delimited by the green contour line), which represents an improvement of at least a factor of five, and up to two orders of magnitude, over the standard MS approach.

Moving to higher motional states ($n=5$), the present scheme demonstrates a more significant advantage. The region with very substantial improvement (in red and yellow tones) expands to encompass larger errors across both axes. This improvement also decays more gradually with higher decay rates compared to the case with motional ground states (first row). Importantly, the MS gate fails to achieve fidelities exceeding $99\%$ for any frequency error within the studied range. In contrast, the robust scheme maintains high fidelities exceeding $99\%$ within a broad range of errors (up to $\epsilon_{\nu} < 3\%$ and $\epsilon_{\omega} < 2\%$) and even achieves exceptional fidelities exceeding $99.9\%$ with frequency errors of any type of up to $0.7\%$. This advantage persists even for high motional occupations ($n\sim 100$), where the robust scheme still maintains fidelities exceeding $99\%$ with moderate frequency errors. 

\section{Methods} \label{methods}

This section details the derivation of the driving protocol in Eq.~\eqref{eq:parametrisation} that implements a robust entangling gate operation. The approach involves first designing the general driving pattern for the desired operation and then refining it to enhance robustness against noise and imperfections.
 
\subsection{Entangling gate} \label{Protocol}

While Eq.~\eqref{eq:hamiltonian_simplified} describes the ion-laser interaction Hamiltonian with a driving pattern $\tilde f_j(t)$ near-resonant with all sideband transitions, only a finite subset of those sidebands is needed for the implementation of the entangling gate. This selection process involves specifying the time-dependent functions $F_{l,k}^{(j)}(t)$ for specific side-bands ($k$) of specific motional modes ($l$). Crucially, achieving a successful implementation relies on both a well-designed set of functions $F_{l,k}^{(j)}(t)$ and an appropriate choice of sidebands. If the chosen set of sidebands is inadequate, there may not exist a modulation strategy (i.e., a choice of $F_{l,k}^{(j)}(t)$) that fulfills both requirements at the gate time: (i) eliminating undesired spin-motion coupling terms and (ii) reaching the desired Rabi angle $\phi_T$ for the entangling term.

Once a sideband set and modulation strategy are chosen, one can analyze the system's evolution.  
An initial spin-motion state $\rho = \rho_S \otimes \rho_M$ evolves under the influence of the time-dependent Hamiltonian $H(t)$ in Eq.~\eqref{eq:hamiltonian_simplified} to a final state $\rho(T) = U(T)\rho U^{\dagger}(T)$, with the unitary evolution operator $U(T) = \mathcal T \exp(-i \int_0^T dt H(t))$. Constructing $U(T)$ for a chosen modulation strategy thus reveals the effective operators present at the gate time and therefore the final state $\rho(T)$. The Magnus expansion \cite{blanes2009magnus} provides an approximation of $U(T)$ using a finite number of terms, effectively acting as an exact solution up to a specific order in the Lamb-Dicke parameter $\eta$. This work employs a third-order solution in $\eta$, surpassing the standard first-order approach, which leads to a more accurate model of the system's dynamics. 

Specifically, the interaction Hamiltonian to order $\eta^3$ is given by

\begin{align}
\label{hamiltonian_eta3}
\nonumber H(t) &=  \sum_{j} \left[ i F_{1,1}^{(j)} \sigma_y^{(j)} a_1^{\dagger} \left( \unit - \frac{1}{2} \eta_{j1}^2  a_1^{\dagger} a_1 - \sum_{l=2}^M \eta_{jl}^2 a_{l}^{\dagger} a_l  \right) \right. \\
& \left. \qquad - \frac{1}{2}  \sum_{l=1}^M \eta_{jl} F_{l, 2}^{(j)}  \sigma_y^{(j)} a_l^{\dagger 2} \right] + h.c. \ ,
\end{align}
with the creation and annihilation operators $a_1^{\dagger}$ and $a_1$ referring to the designated central motional mode. Accordingly, the unitary evolution operator induced by this Hamiltonian can be found analytically with the Magnus expansion truncated to fourth order, \textit{i.e.}, ${U(t) = \exp(\sum_{i=1}^4 \mathfrak H_i(t))}$, where

\begin{subequations}
\begin{align}
\mathfrak H_1 &= -i \int \mathbf{dt}_1\, H_1 \\
\mathfrak H_2 &= \frac{-1}{2} \iint \mathbf{dt}_2\, \comm{H_1}{H_2} \\
\mathfrak H_3 &= \frac{i}{6} \iiint \mathbf{dt}_3\, \Bigl( \comm{H_1}{\comm{H_2}{H_3}}  + \comm{H_3}{\comm{H_2}{H_1}} \Bigr) \\
\nonumber \mathfrak H_4 &= \frac{1}{12} \iiiint \mathbf{dt}_4\, \Bigl( \comm{\comm{\comm{H_1}{H_2}}{H_3}}{H_4} \\
\nonumber &\quad \qquad \qquad \qquad + \comm{H_1}{\comm{\comm{H_2}{H_3}}{H_4}} \\
\nonumber &\quad \qquad \qquad \qquad  + \comm{H_1}{\comm{H_2}{\comm{H_3}{H_4}}} \\
&\quad \qquad \qquad \qquad  + \comm{H_2}{\comm{H_3}{\comm{H_4}{H_1}}} \Bigr) \ , 
\end{align}
\end{subequations}
with the short-hand notations $H_i \equiv H(t_i)$ and $\mathbf{dt}_j\equiv dt_1\cdots dt_j$. This expansion reveals that $U(T)$ becomes an exponential of a sum of operators of the form $g(T) \left(\prod_l a_l^{\dagger p_l} a_l^{q_l}\right) \left(\sigma_y^{(1)}\right)^{r}\left(\sigma_y^{(2)}\right)^{s}$ for some integers $p_l,q_l, r, s$ and function $g(T)$ dependent on the integrals over the driving functions $F_{l,k}^{(j)}(t)$. 

Crucially, not all these terms contribute to the desired entangling operation. To achieve pure entanglement at the gate time $T$, the driving functions $F_{l,k}^{(j)}$ need to be carefully designed to eliminate all terms except for the one responsible for the entangling operation $i \phi_T \sigma_y^{(1)}\sigma_y^{(2)}$. This selective elimination translates to a set of constraints on the driving functions, that have been provided in electronic format for convenience \cite{data_zenodo}.

One can then verify that the monochromatic driving functions 
\begin{align}\label{eq:parametrisation3}
\nonumber F_{1,1}^{(1)} &= F_{1,1}^{(2)} =  \Omega e^{2i \delta t} \\
\nonumber F_{l,2}^{(1)} &=  s_{1l} \Omega  \frac{\tilde \eta_l}{\eta_{1l}} e^{i \delta t} \\
F_{l,2}^{(2)} &= s_{1l} \Omega \frac{\tilde \eta_l }{\eta_{2l}} e^{i \delta t} , 
\end{align}
with $\tilde \eta_l := \frac{\sqrt{2}}{2} \sqrt{\eta_{1l}^2  + \eta_{2l}^2}$ and the entangling condition $-3 \frac{\Omega^4}{\delta^4} \left( \sum_j \eta_{j1}^2 \right) +  \frac{\Omega^2}{\delta^2} \left(  2 +  \sum_{l, j} \eta_{jl} ^2 \right)  = \frac{\phi_T}{\pi}$ are sufficient to satisfy all those constraints and, therefore, to eliminate undesired contributions, as well as to achieve entanglement at the gate time $T$. Notably, while the driving scheme based on Eq.~\eqref{eq:parametrisation3} achieves entanglement, it offers limited robustness compared to the solution presented in Eq.~\eqref{eq:parametrisation}.

\subsection{Robustness to motional heating} \label{robustness}

Motional heating, caused by interaction with the environment, significantly degrades gate fidelity by introducing correlations between qubits and motional degrees of freedom. Thermalisation and dephasing, the leading sources of error, can be modeled by the master equation \cite{breuer2002theory}
\begin{align} \label{eq:master}
\frac{d}{dt} \rho = -i \comm{H(t)}{\rho} + \sum_{l=1}^M \mathcal L^{(l)}(\rho) \ ,
\end{align}
with the density matrix $\rho = \rho_S \otimes \rho_M$, the Hamiltonian $H(t)$ in Eq.~\eqref{eq:hamiltonian_simplified} and the Lindbladian $ \mathcal L^{(l)}(\rho)$  that includes the jump operators $C_{-}^{(l)} = a_l$  and $C_+^{(l)} = a_l^{\dagger}$ for thermalization in the $l$-th mode and $C_n^{(l)} = a_l^{\dagger} a_l$  for dephasing:

\begin{align}
\label{eq:lindblad}
\mathcal L^{(l)}(\rho) = \sum_{j=-, +, n} \gamma_{j}^{(l)} \left( C_j^{(l)} \rho C_j^{(l)\dagger} - \frac{1}{2} \{C_j^{(l)\dagger} C_j^{(l)}, \rho \} \right).
\end{align}
Here, $\gamma_{j}^{(l)}$ are the decay rates associated to the $j$-th operator of the $l$-th motional mode and $\{ \cdot, \cdot\}$ the anticommutator.

In the absence of dissipation, the target gate $U$ is perfectly implemented, \textit{i.e.} $\rho(T) = U \rho(0) U^{\dagger}$, but the presence of dissipating processes $\mathcal L^{(l)}(\rho)$ introduces noise, resulting in an imperfect (potentially non-unitary) gate $\tilde{\mathcal{E}}: \rho \rightarrow \tilde {\mathcal{E}}(\rho)$. Since this distortion depends on the evolution of the motional state, the choice of the driving functions $f(t)$ in $H(t)$ significantly impacts the susceptibility of $U$ to these dissipative effects. The M{\o}lmer-S{\o}rensen gate is particularly vulnerable to motional heating due to its non-centered phase-space trajectories, but alternative choices of the driving functions $F_{l,k}(t)$ can significantly enhance gate's resilience to heating \cite{haddadfarshi2016high}.

In the time-dependent frame defined by the non-dissipative unitary $U(t)$ (defined before Eq.~\eqref{hamiltonian_eta3}), as long as the entangling conditions on the functions $F_{l,k}^{(j)}(t)$ are met, the coherent evolution reduces to the identity $\unit$, and the the master equation in Eq.~\eqref{eq:master} becomes purely dissipative $\frac{d}{dt} \rho = \sum_l \tilde{\mathcal{L}}^{(l)}(\rho)$ with 
\begin{align}
\label{eq:lindblad2}
\tilde{\mathcal{L}}^{(l)}(\rho) = \sum_{j=-, +, n} \gamma_{j}^{(l)} \left( \tilde C_j^{(l)} \rho \tilde C_j^{(l)\dagger} - \frac{1}{2} \{ \tilde C_j^{(l)\dagger} \tilde C_j^{(l)}, \rho \} \right),
\end{align}
where the time-dependent jump operators are now expressed in the rotated frame ${\tilde C_j(t) = U^{\dagger}(t) C_j U(t)}$, and their explicit expressions up to order $\eta^3$ can be found in electronic format in \cite{data_zenodo}.  Given that the decay rates satisfy $\gamma_jT \ll 1$, the dynamics of $\rho(t)$ can be described at first order by the propagator $e^{\tilde{\mathcal{L}}t} \approx \unit + \sum_l \int_0^t dt' \tilde{\mathcal{L}}(t')$. Specifically, the spin state evolves to a final state at the gate time $ \rho_S(T) =  \rho_S(0) + \Delta \rho_S$, with
\begin{align}
   \Delta \rho_S = \tr_M \int_0^T dt' \tilde{\mathcal{L}}(t') [\rho_S \otimes \rho_M] \ , 
   \label{eq:delta_rhos}
\end{align}
and $ \tr_M$ denotes trace over the motion. While in the absence of dissipation Eq.~\eqref{eq:delta_rhos} vanishes, thermalization and dissipation result in a non-zero $\Delta \rho_S$,  which can be generally expressed as
\begin{align} \label{eq:deltarho}
   \Delta \rho_S = \sum_{m,n} \chi_{mn} A_m \rho_S(0) A_n^{\dagger} \ ,
\end{align}
with the operators $A_m \in \{\ \unit, \sigma_{y}^{(1)}, \sigma_{y}^{(2)}, \sigma_{y}^{(1)}\sigma_{y}^{(2)} \}$. Here, $\chi_{mn}$ are the process matrix elements whose norm characterizes the infidelity of the gate and depend on the motional properties of the state through elements of the form $\tr_M \left( \rho_M \prod_l a_l^{\dagger p_l} a_l^{q_l} \right)$ (for some integers $p_l, q_l$) as well as on the temporal shaping of the driving functions $F_{l,k}^{(j)}(t)$. Therefore, since each element $\chi_{mn}$ is weighted with a factor of $\eta$, it is possible to devise schemes that are robust to a particular order of $\eta$. Formulated as an optimisation problem, this entails determining the form of the drivings $F_{l,k}^{(j)}(t)$ that minimize a set of contraints, as shown in Supp. Mat. \ref{Supmat1}. 

While the parametrization in Eq.~\eqref{eq:parametrisation3} offers limited opportunities to improve its robustness, a bichromatic drive with minimal modifications to the protocol of Eq.~\eqref{eq:parametrisation3}, which results in Eq.~\eqref{eq:parametrisation}, can significantly improve robustness to the leading order $\order{\eta^0}$ by minimizing the functionals in Eq.~\eqref{eq:functionals} of the Supp. Mat.~\ref{Supmat1}.

\section{Conclusions}
\label{conclusions}

Achieving high-fidelity entangling gates is a critical bottleneck for trapped-ion quantum computation. Current gate implementations, while often tailored to address specific error sources, struggle when confronted with the combined effects of thermal excitations, control field imperfections, and environmental coupling. These factors lead to unwanted spin-motion couplings, detuning errors, and motional heating, ultimately degrading gate performance. Consequently, high-fidelity operations often require highly controlled environments and sophisticated control techniques, even to achieve moderate entanglement fidelities with small ion chains.

This work presents an entangling gate scheme that tackles the fundamental challenge of mitigating diverse error sources simultaneously. Our proposed gate utilizes a specifically designed driving protocol that addresses multiple sidebands of each motional mode. This approach effectively suppresses inhomogeneous spin-motion couplings, leading to significant improvements in gate fidelity, even for large ion chains with numerous motional modes. Furthermore, the driving patterns enable high-fidelity operations even under substantial thermal excitations, a crucial factor for scalability.

The proposed scheme demonstrates remarkable robustness against motional heating, vibrational frequency fluctuations, and detuning errors. This combined robustness and scalability pave the way for significant advancements in trapped-ion quantum computation. By enabling high-fidelity entanglement under noisy conditions, this work brings us closer to the realization of practical quantum computers.

\section{Acknowledgements}

We gratefully acknowledge stimulating discussions with Jake Lishman, Nguyen Le, Winfried K. Hensinger and Samuel J. Hile. This work was supported by the U.K. Engineering and Physical Sciences Research Council via the EPSRC Hub in Quantum Computing and Simulation (EP/T001062/1).

\bibliography{biblio}

\newpage
\appendix{\bf{SUPPLEMENTARY MATERIAL}}

\section{Functionals for resilience against motional heating}\label{Supmat1}

The gate fidelity $F$ (Eq.~\eqref{eq:gate_fidelity}) in the presence of motional heating depends on the temporal modulation of the driving functions $F_{l, k}^{(j)}$. Specifically, the fidelity $F$ can be maximized by selecting these functions $F_{l, k}^{(j)}$ to minimize the matrix elements $\chi_{m,n}$ in Eq.~\eqref{eq:deltarho}, or equivalently, a set of cost functions $\{ c_i \}$ evaluated at the gate time $t=T$, which naturally arise at various orders in the Lamb-Dicke parameter $\eta$. \newline 

\paragraph*{Order $\eta^1$}
The first-order-in-$\eta$ contributions to the motional heating are functionals of the driving functions corresponding to the central motional mode, $F_{1,1}^{(j)}$, with $j=\{ 1,2 \}$. Specifically, the cost functions read
\begin{align}
\nonumber 
& c_1 = \{ \alpha_i \} \ , && 
c_2 = \{ \alpha_i^2 \} \ , && 
c_3 = \{ \alpha_i \alpha_j \} \ , \\
\nonumber 
& c_4 = \{ |\alpha_i|^2 \alpha_j \} \ , && 
c_5 = \{ \alpha_i^2 \alpha_j^{*2} \} \ , &&
c_6 = \{ |\alpha_i|^2 \} \ , \\ 
& c_7 = \{ \alpha_i \alpha_j^* \}  \ , && 
c_8 = \{ |\alpha_i|^2 |\alpha_j|^2 \} \ , &&
c_9 = \{ \alpha_i^2 \alpha_j^* \} \ ,
\label{eq:functionals}
\end{align}
with $i\neq j$, $\alpha_j :=  \{ F_{1,1}^{(j)} \}$ and with the short-hand notation $\{ f \} = \int_0^t dt' f(t')$. 

Crucially, a Fourier series parametrization of the form $f = \sum_m A_m e^{i a_m t}$ with the constraint $\sum_m \left( A_m \prod_{k\neq m} a_k \right) = 0$ for the functions $F_{1,1}^{(j)}$ ensures that all conditions vanish $c_i=0$, except for $i=\{6,8\}$. Consequently, optimal motional heating resilience at first order in $\eta$ can be achieved through the minimization of the functional $C = \max(c_6 , c_8)$. 
\newline

\paragraph*{Orders $\eta^2$ and $\eta^3$} Second and third-order contributions to motional resilience involve numerous functionals, becoming significant only in the regime of large Lamb-Dicke parameters. For $2$ ions, there are $186$ and $2543$ independent elements for the respective orders, available conveniently in electronic format in \cite{data_zenodo}.

\section{Highly excited thermal states}\label{Supmat2}

While Figure~\ref{fig:2} in the Main text showcases the performance of the proposed gate for various ion chain sizes and scenarios including ideal conditions or motional heating, the results are limited to motional Fock states $\ket{\mathbf{n}}$ with occupation numbers of up to $n=10$. 

\begin{figure}[h!]
     \centering     \includegraphics[width=0.5\textwidth]{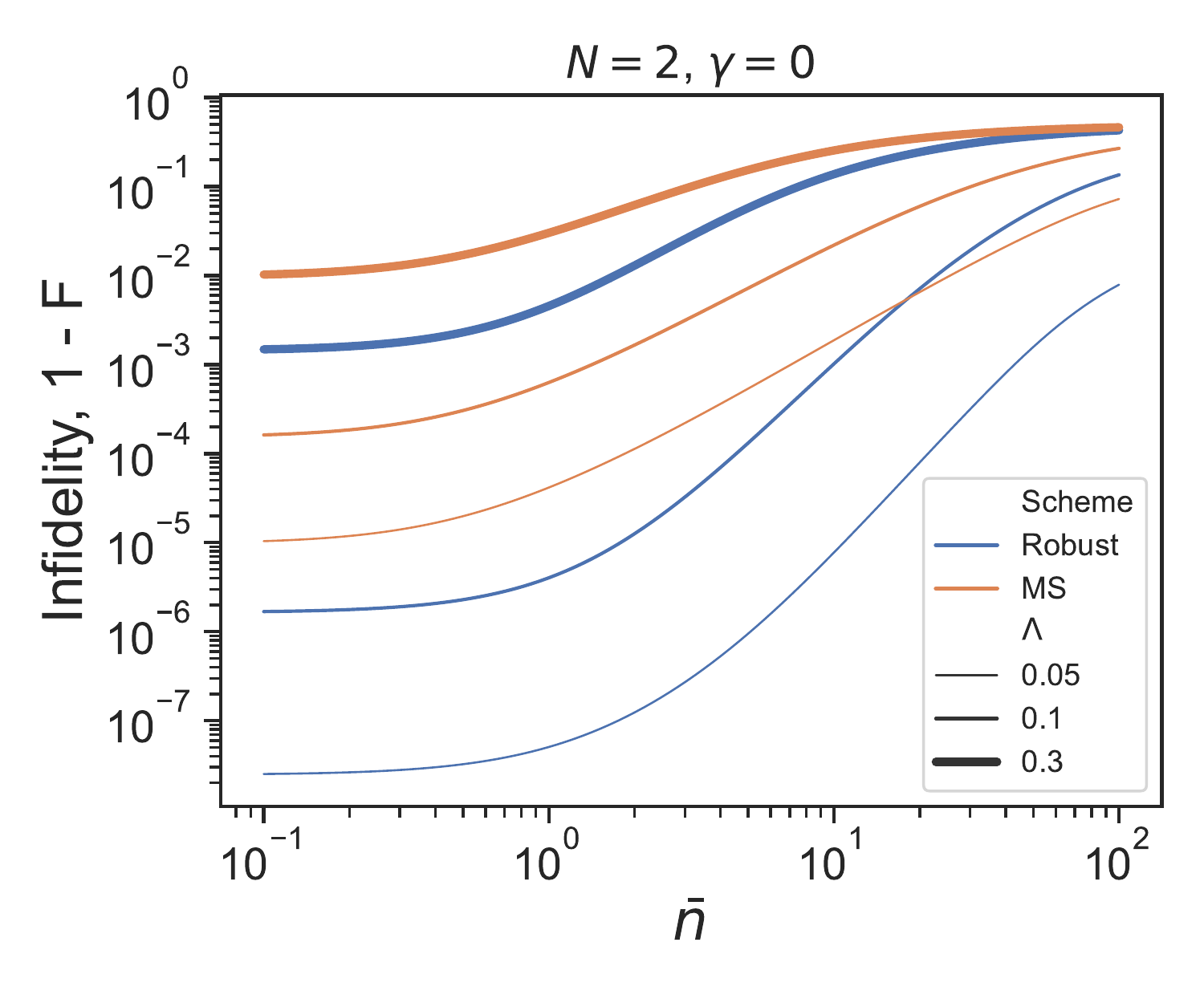}
     \caption{Infidelity $1-F$ for $N=2$ qubits as a function of the mean occupation $\bar n$ of a thermal state (on log scale). Increasing line thickness represents increasing values of the coupling parameter $\Lambda$ (defined in the main text), while the color line distinguishes between the conventional MS approach (orange) and the present robust gate scheme (blue). No motional heating or detuning errors are considered.}
     \label{fig:supp_a}
\end{figure}

Fig.~\ref{fig:supp_a} depicts the infidelity of a two-ion system for thermal states with average phonon occupations (thermal excitation) of up to $\bar n = 100$ and different spin-motion coupling strengths $\Lambda$. The figure demonstrates that the proposed scheme outperforms the standard MS approach for any value of $\bar n$ and $\Lambda$. Remarkably, fidelities exceeding $99\%$ are achieved with spin-motion couplings $\Lambda$ of $0.05$, a value commonly encountered in experiments, which represents more than an order of magnitude improvement compared to the standard MS approach.

\end{document}